# Interpreting Intensity Speckle as the Coherency Matrix of Classical Light

KyeoReh Lee[1,2,*] and YongKeun Park[1,2,3,†1]

[1]*Department of Physics, Korea Advanced Institute of Science and Technology (KAIST), Daejeon 34141, Republic of Korea*
[2]*KAIST Institute for Health Science and Technology, Daejeon 34141, Republic of Korea*
[3]*Tomocube Inc., Daejeon 34109, Republic of Korea*

We show that an intensity speckle can be directly interpreted as the properties of incident light – amplitude, phase, polarization, and coherency over spatial positions. Revisiting the speckle-correlation scattering matrix (SSM) method [Lee and Park, Nat. Comm. 7, 13359 (2016)], we successfully extract the intact information of incident light from an intensity speckle snapshot as the form of coherency matrix. The idea is verified experimentally by introducing the peculiar states of light that exhibit uneven amplitude, phase, polarization, and coherency features. We also find substantial practical advantage of the proposed method compared to the conventional coherency matrix measuring techniques such as Stokes polarimetry. We believe this physical interpretation of an intensity speckle could open a new avenue to study and to utilize the speckle phenomenon in vast subfields of wave physics.

## I. INTRODUCTION

Speckle patterns are the characteristic granular structure that is commonly observed in a coherent system in both unintentional and intentional manners [1,2]. In imaging systems, such granularity has been treated as a noise that significantly reduces image contrast [3-5]; but in speckle metrology, on the other hand, it has been utilized to detect the minute variations in the frequency and wavefront of light [6-12].

Speckle is the consequence of interference that is generally observed when coherent waves pass through complex media such as ground glasses, rough surfaces, and biological tissues, which introduce disordered spatial phase variations [2]. Though it seems arbitrary, the formation of speckle grains is not a random or stochastic process, but a deterministic process. It is predictable if input light and a diffusive optical system are known. In Ref. [13], Popoff et al. demonstrated the fact by reconstructing the initial optical field from the speckle field, exploiting an optical transmission matrix. They showed that optical diffusers are not fundamentally different from conventional optics, but are the same linear systems having more complex transmission (or reflection) matrices [14]. Accordingly, optical diffusers convey the optical information as in conventional optical systems, but in more scrambled forms – the speckle fields.

One other intriguing feature of speckles is their statistical property. Despite the numerous different situations of speckle generation, the speckle fields mostly show complex Gaussian distribution, whose amplitudes follow the Rayleigh distribution [2]. Thus, though it is possible to tailor the odd speckles that follow non-Gaussian statistics using the transmission matrix, immediate return to the Gaussian distribution was observed even by the slight defocus of detection plane [15]. Such solid statistical property of speckle enables powerful mathematical tools of Gaussian random variables, which is especially useful in the intensity speckle analyses. A representative example is the Siegert relation that connects the intensity and field correlation of speckles, which has been routinely utilized in dynamic light scattering analyses [16,17], and is also the underneath principle of speckle metrology techniques [18]. However, despite the advantages and usages of intensity speckles, the physical interpretation of such feasibility has not been clear. What does the intensity speckle represent out of incident light; in other words, how much optical information could be encoded as an intensity speckle, and be retrieved from it? In this letter, we seek an answer to the question.

In order to embrace the deterministic and statistical natures of speckle, we revisit the speckle-correlation scattering matrix (SSM) proposed recently [19-21]. The SSM method is a novel transmission-matrix-based holographic technique. Unlike the previous works that had presented linear inversion of transmission matrix from the complex speckle *field* [13,14,22,23], the SSM provides the solid way of achieving the complex incident field from a single *intensity* speckle without reference wave, additional constraints, nor multiple measurements [19].

Here, we extend the idea of SSM to the more general properties of the classical light. We show that the SSM is a covariance matrix over a space-polarization domain that has been used to define the general state of light, including mixed (or partially coherent) states called coherency matrix [24,25]. Since the SSM is calculated from an intensity speckle without any additional information (notice the transmission matrix is the predetermined constant of a given optical system), we find it is an intensity speckle that determines the entire coherency matrix of light. We also discuss the theoretical and practical constraints of the proposed idea.

---

[*]*kyeo@kaist.ac.kr*

[†]*yk.park@kaist.ac.kr*

The proposed idea is demonstrated experimentally by introducing the general state of light throughout the polarization and mixed states. We also find substantial advantages of the proposed method on the characterization of light, compared to the conventional Stokes polarimetry.

## II. PRINCIPLE

Let us consider a simple optical setup for intensity speckle measurement (Fig. 1). Assume that the speckle is fully developed, and exhibits the Gaussian statistics on the detection plane. Inappropriate selection of a diffuser or a propagation distance between the diffuser and a detector may result in the formation of non-Gaussian speckle [26,27]. Once the setup is fixed, the relationship between incident and speckle fields is firmly determined by the transmission operator, $\hat{t}$ [14,28,29].

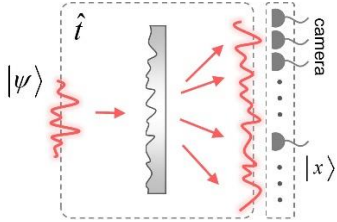

FIG. 1. Intensity speckle measurement. A diffuser transforms the incident light into speckle field. A proper diffuser and a propagation length need to be chosen to ensure the complex Gaussian probability distribution of speckle along the $|x\rangle$ space.

Now suppose an incident light of pure state $|\psi\rangle$, which is coherent light having well-defined (static) wavefront and polarization state. It could, therefore, be characterized as a complex-numbered vector, $\psi_e = \langle e|\psi\rangle$, by taking a certain basis $\sum_{e=1}^{N} |e\rangle\langle e| = 1$ of the position-polarization Hilbert space that $|\psi\rangle$ spans. The acquired intensity speckle could be described as $I_x = \beta_x^* \beta_x$, where $\beta_x = \langle x|\hat{t}|\psi\rangle$ is the generated speckle field, while $|x\rangle$ represents the lateral position over the detection plane. Then, the transmission matrix could also be embodied as $t_{xe} = \langle x|\hat{t}|e\rangle$. Note that $t_{xe}$ is also another speckle field over the detection plane for $|\psi\rangle = |e\rangle$. At this point, we introduce the SSM defined by

$$Z_{ij} = \frac{1}{\sigma_i \sigma_j}\left[\left\langle t_{xi}^* t_{xj} I_x \right\rangle_x - \left\langle t_{xi}^* t_{xj} \right\rangle_x \left\langle I_x \right\rangle_x\right], \quad (1)$$

where $\langle \ \rangle_x = \frac{1}{M}\sum_{x=1}^{M}$ and $\sigma_e = \left\langle t_{xe}^* t_{xe} \right\rangle_x$ [19]. Utilizing the Gaussian statistics of speckles, the first term of Eq. (1) could be decomposed by the Wick's (or Isserlis') theorem that holds for Gaussian random variables [30],

$$Z_{ij} = \frac{1}{\sigma_i \sigma_j}\left[\left\langle t_{xi}^* \beta_x \right\rangle_x \left\langle t_{xj} \beta_x^* \right\rangle_x + \left\langle t_{xi}^* \beta_x^* \right\rangle_x \left\langle t_{xj} \beta_x \right\rangle_x\right]. \quad (2)$$

When the number of samplings is large enough to satisfy $M \gg N$, the general orthogonality relations $\sqrt{\frac{1}{\sigma_i \sigma_j}}\left\langle t_{xi}^* t_{xj} \right\rangle_x = \delta_{ij}$ and $\left\langle t_{xi} t_{xj} \right\rangle_x = 0$ holds in speckle [2]. Then, the first term in Eq. (2) only survives and becomes

$$Z_{ij} = \langle e = i|\psi\rangle\langle\psi|e = j\rangle, \quad (3)$$

which is the shape of density matrix (in quantum regime) of the incident light $\hat{\rho} = |\psi\rangle\langle\psi|$ in a given basis $\sum_{e=1}^{N} |e\rangle\langle e| = 1$ [31]. In classical coherence theory, such covariance matrix has also been utilized to determine the general coherency properties between position and polarization states, called the cross-spectral density (CSD) matrix [24,32], and coherency matrix [24,25], respectively. Although both terminologies may not be adequate for the Eq. (3) (because it spans position-polarization space at the same time), we decide to use 'coherency matrix' throughout the Letter according to several recent works measured two-position coherency matrix [33-35].

We also found that Eqs. (1-3) are valid in the mixed (or partially coherent) states of light, which composed of more than one microstates that are incoherent to each other. Such classical mixed state could be introduced by imposing independent fluctuations on each microstate. For the mixed state, each microstate $|\psi_\alpha\rangle$ generates its own intensity speckle $\beta_{x,\alpha}^* \beta_{x,\alpha}$, which is incoherently summed over the detection plane. Accordingly, the compounded intensity pattern $I_x = \sum_\alpha P_\alpha \beta_{x,\alpha}^* \beta_{x,\alpha}$ is acquired, where $P_\alpha$ denotes the statistical probability of $|\psi_\alpha\rangle$. Substituting the intensity pattern to the Eq. (1), one can readily find the corresponding SSM also presents the coherency matrix of the incident mixed states of light, $Z_{ij} = \sum_\alpha P_\alpha \langle e=i|\psi_\alpha\rangle\langle\psi_\alpha|e=j\rangle$. This result implies that a simple optical diffuser in front of the camera is just enough to read the complete optical coherency information of the light over position-polarization space. However, before we proceed, several theoretical and practical constraints should be explored.

First, the speckles $\beta_x$ or $t_{xe}$ can be non-Gaussian. The optical diffuser should be selected and placed carefully to ensure the Gaussian statistics of speckle on the detection plane (Fig. 1).
Second, the speckles from the different basis vectors may be significantly correlated $\left\langle t_{xi}^* t_{xj} \right\rangle_x \neq 0$ for $i \neq j$, which disables the distinguishability between the two different basis vectors and effectively reduce the dimension of measurable Hilbert space. We find such distinguishability is related to the property of optical diffuser. For example, intensity speckles generated by the commercial ground glasses are usually insensitive to the polarization of light, while multiple scattering media are much more sensitive to it [36,37]. In other words, the sensitivity of intensity speckle determines the measurable domain. As several previous works have already proved the sensitivity of intensity speckle to the various properties of light [38-40], we expect the proper selection of diffusers can readily achieve the desired degree of measuring capability. Since multiple scattering usually increases such sensitivity, introducing a scattering medium as a diffuser can be a good choice for the sake of distinguishability. However, it is noteworthy that the multiple scattering usually decreases the transmittance as a trade-off, which may induce the correlation between different speckles [41-45].

Third, as analyzed in our previous work [19], the oversampling ratio (or $M/N$), $\gamma$ plays an important role in noise handling. It was shown that higher $\gamma$ is required for stable results as the practical noise level increases. Even in ideal situations without any noise, $\gamma$ should be larger than 4 for the safe reconstruction of the pure state due to the inherent noise (the second) term in Eq. (2). For mixed states, we will see the required $\gamma$ also increases linearly with the number of microstates in a mixed state, or $\text{rank}(\hat{Z})$, where $\hat{Z}$ is an operator representation of coherency matrix. Please notice we utilize an additional error reduction algorithm to rule out such noise effect (see Appendix A). Note that the generality of our method is still valid since the additional sequence does not require any additional information or free variable as in regularization-based estimation methods [46].

## III. EXPERIMENTAL RESULTS

In order to demonstrate the proposed idea, we prepare a compact unit assembled with a camera [Fig. 2(a)]. The unit is composed of only three optical components – an iris, a diffuser, and a polarizer. The iris is added before the diffuser to block ambient light, and the polarizer is inserted to fix the detection polarization state. The distance between the diffuser and the camera is adjusted to assure that the size of optical modes (i.e., speckle gain) over the detection plane is larger than the camera pixel size.

A rutile diffuser was introduced in order to maximize the birefringence without sacrificing light transmissivity. We deposited rutile nanoparticles (particle size $\leq 100$ nm) on both sides of a coverslip using a spray painting method [Fig. 2(b)].

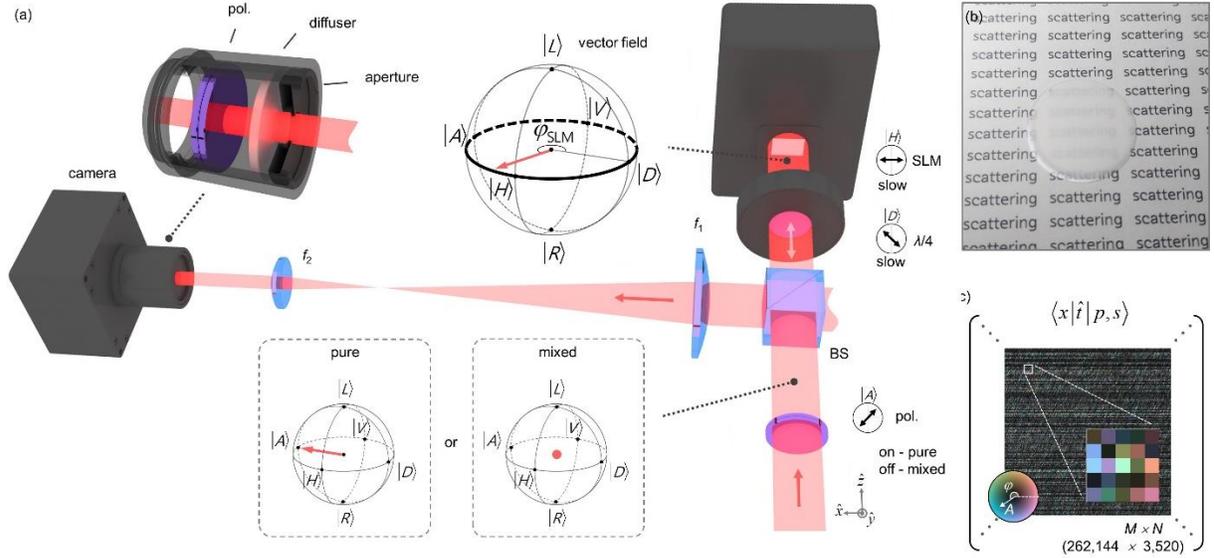

FIG. 2. Experimental setup. (a) The optical setup for preparing and measuring the state of light. Existence of the polarizer in front of the laser determines the prepared polarization state to either pure $|A\rangle = \frac{1}{\sqrt{2}}|H\rangle - \frac{1}{\sqrt{2}}|V\rangle$ or mixed $\frac{1}{2}|H\rangle\langle H| + \frac{1}{2}|V\rangle\langle V|$ state. In case of the pure state, the output polarization state is a function of $\varphi_{SLM}$, which varies over the SLM active area. In the case of the mixed state, the SLM phase pattern only modulates the $|s = R\rangle$ state. (b) The custom-made diffuser made of rutile nanoparticles. The rutile nanoparticles were deposited on the 25 mm diameter coverslip by a spray painting method. (c) Calibrated transmission matrix $t_{xps}$ of the rutile diffuser. Within the total $M$-by-$N$ transmission matrix (262,144 × 3,520), the central 512 × 512 subpart is displayed. In the color circle, the $A$ and $\varphi$ denote the amplitude and phase of the complex value in arbitrary and radian units, respectively.

The manufactured diffuser has rutile scattering media thickness of 30 μm on both side of the coverslip (140 μm thickness), mean transmittance of 52% ($\bar{T} = 0.52$), and the transport mean free path of 44.5 μm (see Appendix B).

Due to high birefringence ([47], $\Delta n = 0.288$ for 633 nm) and multiple scatterings, we could achieve the complete decorrelation between the speckles from orthogonal polarizations. Accounting the aperture in front of the diffuser (5.33 × 4 mm, $A = 21.3$ mm$^2$), the dimensionless conductance of the diffuser can be calculated as $g \approx N_0 \bar{T} = 1.77 \times 10^8$, where $N_0 = 2\pi A/\lambda^2$ is the total number of possible input optical modes. Since $g >> M$, the mesoscopic correlation is not significantly observed in this work [48].

The transmission matrix of the diffuser is calibrated using a spatial light modulator (SLM), a He-Ne laser, and a Michelson type interferometry [49] (see Appendix C). We select 44 × 40 plane waves $|p\rangle$ in both horizontal $|H\rangle$ and vertical polarizations $|V\rangle$ as an input basis, $|e\rangle = |p,s\rangle$ ($N$ = 3,520), and 512 × 512 central camera pixels $|x\rangle$ as sampling points ($M' = 262,144$). Thereby, the transmission matrix $t_{xps}$ becomes a 3,520 × 262,144 complex-valued matrix [Fig. 2(c)]. According to Eq. (3), the coherency matrices will be presented in the selected basis, and the optical information spans outside the basis would be considered as noise. Due to the practical oversampling of the intensity speckle (i.e., speckle grain > pixel size), the actual number of sampled optical modes $M$ could be smaller than the number of pixels, $M' \geq M$. We figure out that $M = 49,600$ and $\gamma = 14$ in our experimental scheme by the power spectrum analysis of the acquired intensity speckles (see Appendix D).

### A. Pure states

For the first demonstration, we prepare the pure states using a polarized laser source [Fig. 2(a)]. In order to test the feasibility of the proposed idea, we generate 'vector beams' whose polarizations vary over the transversal position $|r\rangle$ [50].

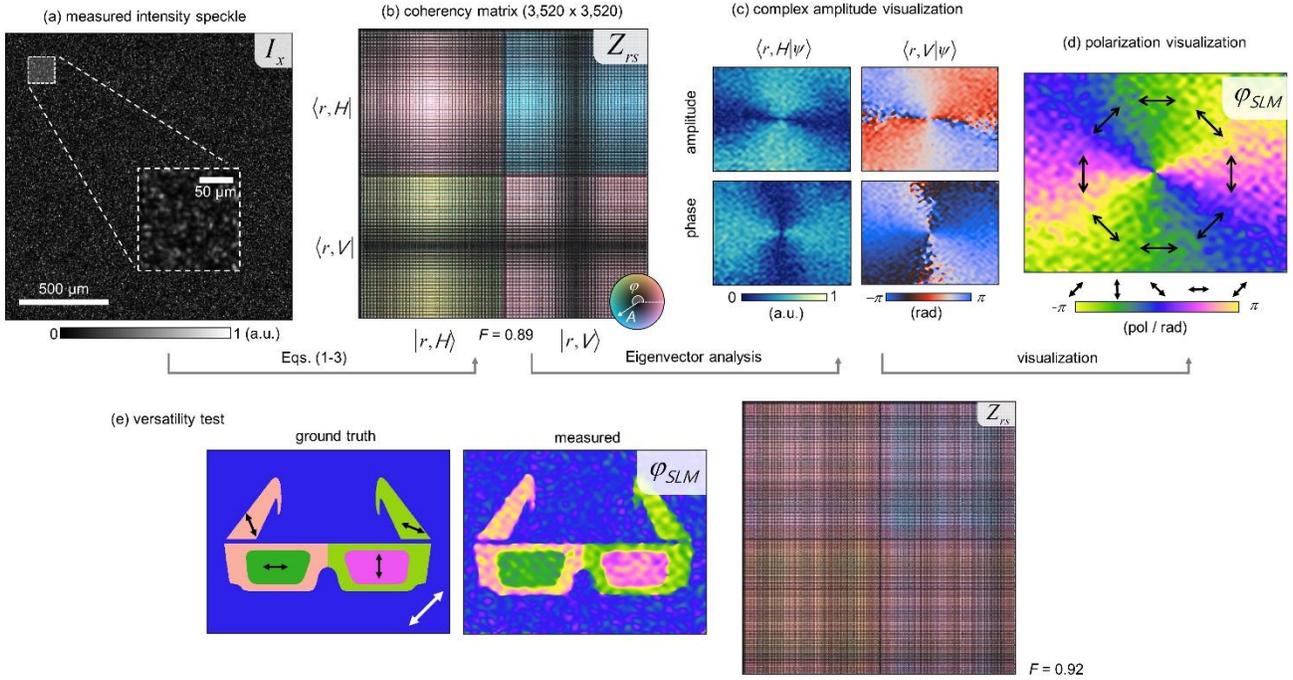

FIG. 3. Experimental demonstrations in pure states. (a) The measured intensity speckle $I_x$. (b) Retrieved $N$-by-$N$ (3,520 × 3,520) coherency matrix $Z_{rs}$ based on the proposed method. In the color circle, the $A$ and $\varphi$ denote the amplitude and phase of the complex value in arbitrary and radian units, respectively. (c) The complex coefficients $\langle r,s|\psi\rangle$ extracted from the coherency matrix result. (d) Visualization of the polarization variation over $|r\rangle$, which is related to the displayed SLM phase pattern $\varphi_{SLM}$. (e) The measured coherency matrix (bottom) in pure state, and the corresponding $\varphi_{SLM}$ patterns (top right) compared with the prepared $\varphi_{SLM}$ (top left). The fidelity values ($F$) between the expected and measured coherency matrices are denoted.

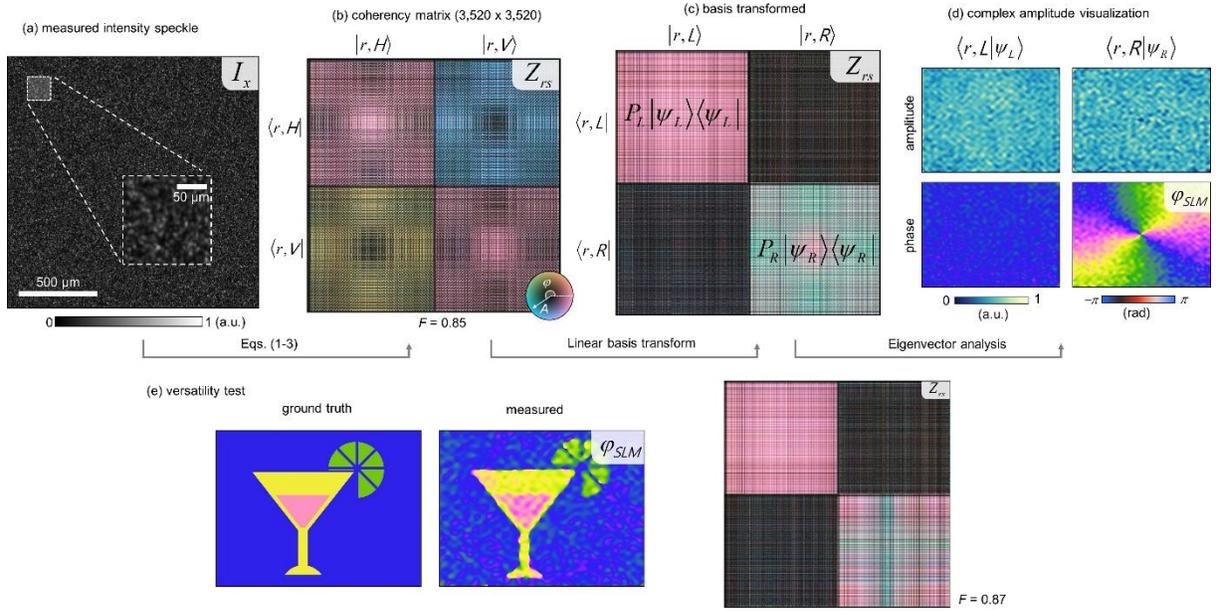

FIG. 4. Experimental demonstrations in mixed states. (a) The measured intensity speckle $I_x$. (b) Retrieved $N$-by-$N$ (3,520 × 3,520) coherency matrix $Z_{rs}$, based on the proposed method represented in the $|H\rangle\langle H|+|V\rangle\langle V|$ polarization basis. In the color circle, the $A$ and $\varphi$ denote the amplitude and phase of the complex value in arbitrary and radian units, respectively. (c) Identical results in $|L\rangle\langle L|+|R\rangle\langle R|$ polarization basis. Here, $P_L = 0.48$ and $P_R = 0.52$ are the measured statistical probabilities of $|\psi_L\rangle$ and $|\psi_R\rangle$, respectively. (d) The two microstates $\langle r|\psi_L\rangle$ and $\langle r|\psi_R\rangle$ cropped out from the coherency matrix is in $|L\rangle$ and $|R\rangle$ polarizations, respectively. Note that only the phase part of the $\langle r|\psi_R\rangle$ is modulated by the SLM. (e) The measured coherency matrix (bottom) in mixed states, and the corresponding $\varphi_{SLM}$ patterns (top right) compared with the prepared $\varphi_{SLM}$ (top left). The fidelity values ($F$) between the expected and measured coherency matrices are denoted.

Utilizing a SLM and a quarter-wave plate, we make the polarizations be the function of a given SLM phase $\varphi_{SLM}$ [Fig. 2(a), Appendix E]. Then, we measure the intensity speckle $I_x$ [Fig. 3(a)], and build the SSM using Eq. (1) and calibrated $t_{xps}$. The detailed procedures for SSM calculation can be found in Ref. [19], and Code 1 [51]. The reconstructed coherency matrices are depicted in Fig. 3(b) For more intuitive visualizations, we used a transversal position $|r,s\rangle$ basis rather than the original plane-wave $|p,s\rangle$ basis by the Fourier transform. By the eigenvector decomposition of the measured coherency matrix, the complex coefficients $\langle r,s|\psi\rangle$ [Fig. 3(c)] and corresponding $\varphi_{SLM}$ [Fig. 3(d)] are retrieved according to Eqs. (3) and (E1), respectively. Intended azimuthal polarization vector beam is well reconstructed as shown in Fig. 3(d). We calculate the fidelity $F = |\langle\psi|\psi_0\rangle|^2$ between expected $|\psi_0\rangle$ and measured $|\psi\rangle$ states to quantify the agreement. We observed consistent high fidelity values, $F \approx 0.9$, regardless of displayed SLM patterns [Fig. 3(e)].

## B. Mixed states

As the next demonstration, we prepare mixed states. While the overall experimental setup is preserved, the light source is converted to an unpolarized laser [Fig. 2(a)].

We confirm the unpolarized state of light by measuring the Stokes parameters (see Appendix F). Since the SLM with a quarter-wave plate modulates $|R\rangle = \frac{1}{\sqrt{2}}|H\rangle - \frac{1}{\sqrt{2}}i|V\rangle$ polarization state only, the prepared mixed states become $\hat{Z}_0 = \frac{1}{2}|\psi_L\rangle\langle\psi_L| + \frac{1}{2}|\psi_R\rangle\langle\psi_R|$, where $|\psi_L\rangle$ and $|\psi_R\rangle$ are in two orthonormal polarization states $|L\rangle$ and $|R\rangle$, respectively. Following the same procedures as in pure states, the coherency matrix can be reconstructed by measuring the $I_x$ [Figs. 4(a-b)]. One simple way to check the validity of the experimental results is to calculate the off-diagonal terms in the circular polarization basis, which are supposed to be zero for the prepared states.

As expected, we find the incoherency in Fig. 4(c) by the basis transformation in polarization space. Further, by cropping out the submatrices, we quantify the statistical probabilities $P_L$ and $P_R$, and the complex coefficients, $\langle r|\psi_L\rangle$ and $\langle r|\psi_R\rangle$ of each microstate [Fig. 4(d)]. Note that only the phase of the $\langle r|\psi_R\rangle$ is modulated by the SLM. We find measured statistical probabilities consistently shows $P_L = 0.48$ and $P_R = 0.52$ within 0.03 standard deviation, which are slightly different from the expectation. Since the optics-derived birefringence or optical activity of the system were calibrated before the measurements using the resting state of SLM (i.e., zero image), we infer such differences are due to the slight reflectivity changes in SLM as the function of an applied voltage.

Again, we calculated the fidelity $F = \left(\text{tr}\left[\sqrt{\sqrt{\hat{Z}_0}\hat{Z}\sqrt{\hat{Z}_0}}\right]\right)^2$ between prepared and measured mixed states, $\hat{Z}_0$ and $\hat{Z}$, respectively, to quantify the agreements as proposed in Ref. [52]. We also observed consistent fidelity values, $F \approx 0.85$, regardless of displayed SLM patterns [Fig. 4(e)]. The lower fidelity values are constantly observed in the mixed state.

In order to elucidate the lower fidelity values in the mixed state, numerical simulations are performed with mixed states having different $\text{rank}(\hat{Z})$ in various $\gamma$. The transmission matrix and complex state of light are arbitrarily generated, but their statistical probability set uniformly to avoid an effective reduction of $\text{rank}(\hat{Z})$ from the inhomogeneous probability assignment. From the generated intensity speckle, the reconstruction of the coherency matrix is done through the proposed way, and the fidelity of the reconstructed field is calculated to quantify the performance, as we show in Code 1 [51]. We find the required oversampling ratio $\gamma_c$ (for $F > 0.95$) linearly increases with $\text{rank}(\hat{Z})$ as $\gamma_c \approx 4 \cdot \text{rank}(\hat{Z})$ even in noise-free circumstances (Fig. 5), which supports our experimental results. As expected, practical noises could further increase the $\gamma_c$, while the positive correlation property to $\text{rank}(\hat{Z})$ remains still. Notice for optical systems having $g \approx M$, the mean transmittance and the condition number of the transmission matrix may also affect the achievable fidelity (see Appendix G).

Until today, Stokes polarimetry has been a conventional way to measure the general state and degree of polarization [53]. It consists of four phase-shifting measurements in a given spatial mode, and the measured Stokes vector is directly related to the four elements of a coherency matrix, $\langle r_0, H|\hat{Z}|r_0, H\rangle$, $\langle r_0, H|\hat{Z}|r_0, V\rangle$, $\langle r_0, V|\hat{Z}|r_0, H\rangle$, and $\langle r_0, V|\hat{Z}|r_0, V\rangle$, which is also equal to common 2 × 2 coherency matrix [25]. We find the same strategy could be extended to measure general $N \times N$ coherency matrix; but, it now requires at least 4 measurements per each $\langle r_1, s_1|$ and $|r_2, s_2\rangle$ pairs. This is not impossible, but going to be very tedious especially due to the scanning spatial mode pairs. For example, our results ($N = 3,520$) could be retrieved by the conventional Stokes polarimetry with $2N(N-1) = 24,773,760$ individual measurements. Therefore, we expect that the proposed method may take a practical advantage in the characterization of the general multimode state of light, due to the single-shot nature.

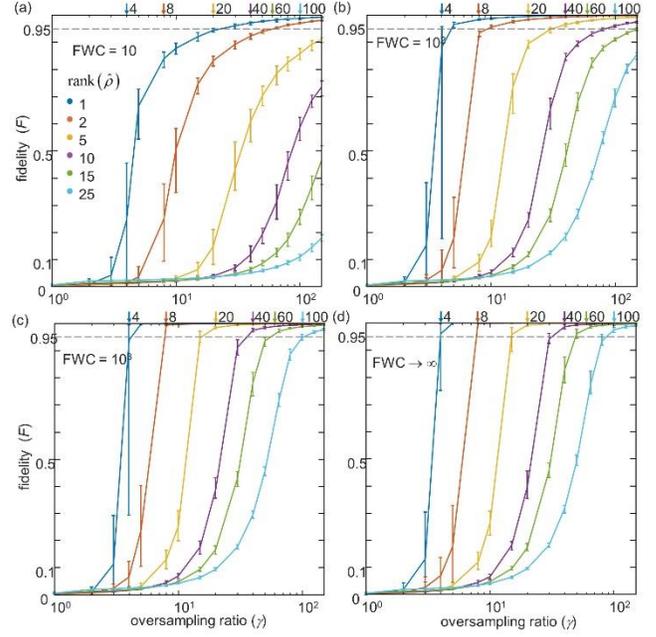

FIG. 5. Numerical fidelity results in shot-noise limited situations. The practical shot-noise is introduced by the finite full well capacity (FWC) of camera. For a given FWC, the shot-noise-limited signal-to-noise becomes $\sqrt{\text{FWC}}$ for the brightest pixels. All fidelity results are calculated after the error reduction algorithm. The error bars represent the 95% confidence intervals of 25 numerical results. Downward arrows above indicate $\gamma = 4 \cdot \text{rank}(\hat{Z})$.

## IV. CONCLUSION

In conclusion, we show that the SSM could be interpreted as the coherency matrix of classical light. Since the transmission matrix is a constant that predetermined by the prepared diffusive optical system, we could deduce that the coherency matrix of light is entirely presented by the measured intensity speckle. The idea is experimentally verified in the general state of light throughout the polarization and mixed states. The required oversampling ratio is numerically explored in a different number of microstates. Compared with the conventional Stokes polarimetry, we have found that the proposed idea could have a substantial advantage in the simultaneous multimodal determination of the coherency matrix.

We expect the present method will open new approaches for the study of wave physics and its application to various disciplines. Because the formation of speckle is fundamentally governed by the wave equations and ubiquitous in various subfields of wave physics, we expect this idea could be generally expanded to the speckles made of different waves such as ultrasound [54], microwave [55], and X-ray [56].

Yet, for the wider utilizations of intensity speckles as an routine light anlysis tool, the stability of a diffusive system, and the reconstruction speed and robustness of coherency matrix should be further improved. We expect the issues can be remedied by the introduction of the engineered diffusers [21,57,58], and advanced non-linear equation solvers, respectively.

## ACKNOWLEDGMENTS

We would like to thank Kitak Kim and Wonjune Choi (Department of Physics, University of Toronto) for the fruitful discussions and comments. This work was supported by KAIST, BK21+ program, National Research Foundation of Korea (NRF) (2017M3C1A3013923, 2015R1A3A2066550, 2018K000396, 2018R1A6A3A01011043)

## APPENDIX A: ERROR REDUCTION ITERATIVE ALGORITHM

To suppress fundamental errors and practical noises, we utilized the Gerchberg-Saxton (GS) type iterative algorithm, while the Fourier transform in the original GS algorithm was replaced by the transmission matrix $t_{xe}$. For each iteration step, the algorithm consists of three sub-steps. First, the basis of the coherency matrix is transformed from $|e\rangle$ to $|x\rangle$ (camera pixel basis) by the transmission matrix, $Z_x = t_{xe} Z_e t_{ex}^+$, where $t_{ex}^+$ is the pseudoinverse matrix of $t_{xe}$ that satisfies $t_{ex}^+ t_{xe} = 1_e$. Second, $Z_x$ is updated to $Z_x^{\bullet}$ by utilizing the measured intensity speckle $I_x$ as a constraint. The amplitude part of $Z_x$ is revised, while the phase part is conserved. Third, the coherency matrix is updated by the inverse basis transform, $Z_e^{\bullet} = t_{ex}^+ Z_x^{\bullet} t_{xe}$. The iteration starts from the Eq. (3), and stops when the coherency matrix converges; the correlation between $Z_e$ and $Z_e^{\bullet}$ reaches 0.999998. In conclusion, the iterative algorithm reduces error by converging the closest local minimum of a given intensity image $I_x$ using the Eq. (3) as an initial guess. Note that the proposed iterative algorithm does not require any additional information or free variables.

## APPENDIX B: CUSTOM RUTILE DIFFUSER

The custom rutile paint was made by mixing rutile nanoparticles (637262, Sigma-Aldrich Co. LLC.) with resin (RSN0806, DOW CORNING®, $n \approx 1.5$ ) and solvent (toluene, 99%) in a proper ratio (0.5 g : 1 mL : 10 mL). In order to disperse the rutile nanoparticles, we sonicated the paint for 10 minutes. The rutile paint was deposited on both sides of the 25 mm diameter and 140 μm thickness round coverslip by the spray painting method using a commercial airbrush (DH-125, Sparmax). After the spray painting, the diffuser was baked (100°C, 10 min) to cure the resin. The diffuser has a thickness of 30 μm per side, mean transmittance of 52% ($\bar{T} = 0.52$), and the transport mean free path of 44.5 μm. The transmissivity and transport mean free path of the diffuser is measured by the integrating sphere (UPK-100-F, Gigahertz-Optik) using the inverse adding-doubling method.

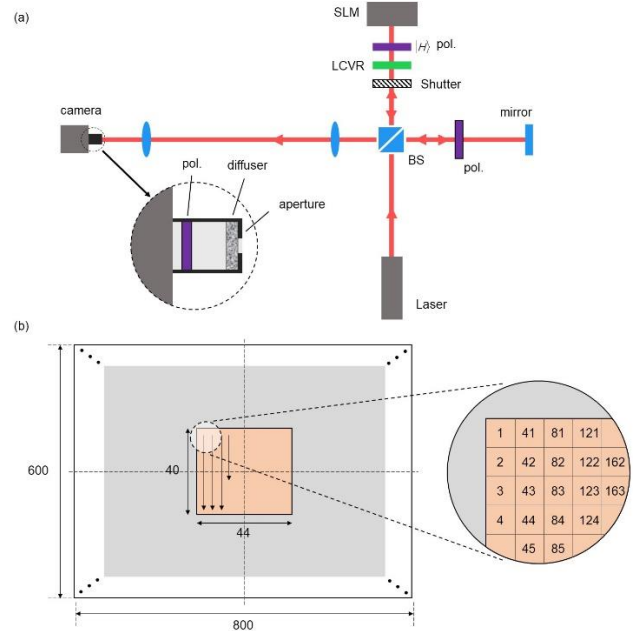

FIG. 6. (a) Optical setup for transmission matrix calibration. The definitions of texts are followings: Laser, unpolarized He-Ne laser (HNL050R, Thorlabs Inc.); BS, 50:50 beam splitter (BS013, Thorlabs Inc.); Shutter, motorized shutter (MFF101/M, Thorlabs Inc.); LCVR, liquid crystal variable retarder (LCC1222-A, Thorlabs Inc.); Pol., polarizer (WP25M-VIS, Thorlabs Inc.); SLM, spatial light modulator (X10468-01, Hamamatsu Photonics K.K.); Camera, CCD camera (MD120MU-SY, XIMEA GmbH). Elements without texts are conventional mirrors and lenses. The 4-$f$ relaying system has ×3 demagnification factor. The LCVR switches the output polarization state. The shutter is used to block the SLM arm in order to measure the reference speckle pattern. The role of the polarizer on the reference arm is to make reference state; therefore, the orientation of the polarizer on the reference arm is not important. (b) Calibrated momentum states, and its vectorization numbering order. The rectangular (800 × 600) momentum space of SLM active area is shown (left). Note that the spacing between the adjacent momentum states is inversely proportional to the clear aperture of the SLM (16 mm × 12 mm). The direct current (DC, or zero transversal momentum) state is placed on the center of the rectangle. The calibrated momentum states are the central 44 × 40 rectangle shape, and the ordering is conventional vectorization order of MATLAB® (right).

## APPENDIX C: TRANSMISSION MATRIX CALIBRATION

Figure 6(a) shows the experimental setup for the transmission matrix calibration. The transmission matrix $t$ is the collection of the speckle field for the specific input states $|\psi\rangle = |p,s\rangle$ measured in the $|x\rangle$ space. Therefore, we produced the $|p,s\rangle$ states using a SLM (X10468-01, Hamamatsu photonics K.K.) and a liquid crystal variable retarder (LCVR; LCC1222-A, Thorlabs Inc.).

The plane waves $|p\rangle$ were prepared by displaying phase ramps on the SLM. We selected a 44 × 40 central area of the reciprocal domain of the SLM. The allocated transversal momentum for the $|p=n\rangle$ state was the $n$-th component of a vectorized 44 × 40 matrix [Fig. 6(b)]. The polarization state $|s\rangle$ was prepared by the LCVR by adjusting the phase retardance.

Since our SLM only modulates the $|s=H\rangle$ state, we achieved output polarization states $|H\rangle$ and $|V\rangle$ for 0 and $\pi$ phase retardations, respectively. Therefore, total 44 × 40 × 2 $|p,s\rangle$ states ($N = 3{,}520$) were prepared.

In order to measure the complex numbers $t_{xps}$, we constructed a Michelson type interferometer using a He-Ne laser (HNL050R, Thorlabs Inc.). The reference arm also supplied a static reference speckle field over the detection plane, $R_x = \langle x|\hat{t}|\text{ref}\rangle$.

Using the phase-shifting concept, we achieved the interference term $R_x^* t_{xps}$. For each $|p,s\rangle$ input state, we took three images $I_{x,0}$, $I_{x,1}$, and $I_{x,2}$, while the SLM displayed $|p,s\rangle$, $|p,s\rangle e^{i\frac{2\pi}{3}}$, and $|p,s\rangle e^{-i\frac{2\pi}{3}}$, respectively. Then, the interference term can be calculated as follows: $R_x^* t_{xps} = \frac{1}{3}\left[I_{x,0} + I_{x,1} e^{-i\frac{2\pi}{3}} + I_{x,2} e^{i\frac{2\pi}{3}}\right]$. The additional $R_x^*$ term must be considered during the calculation of the coherency matrices. Fortunately, we found the phase part of $R_x^*$ would be automatically erased out when we constructed the SSM in Eq. (1). On the other hand, the amplitude part of $R_x^*$ remained as a form of $|R_x|^2$, but was compensated by one additional measurement while the sample arm was blocked. As a result, we took a total of $3N+1$ (10,561) images to calibrate the scattering matrix; it requires about 19 minutes.

We took the central 512 × 512 pixels ($M = 262{,}144$) of the CCD camera (4,242 × 2,830 pixels; 3.1 mm pitch; MD120MU-SY, XIMEA GmbH) for position states $|x\rangle$. Although a greater number of sampling position states provides better results by suppressing the second term in Eq. (2), we could not fully utilize all of the CCD pixels because of limited data storage capacity, and the computing ability of the used computer (3.50 GHz, intel® core™ i5-4690 CPU; 32.0 GB RAM).

The stability of the calibration system [Fig. 6(a)] during the calibration process is confirmed by comparing the speckle patterns before and after the calibration for the same SLM pattern. We observed the calibrated TM remain stable and valid 24 hours after the calibration.

## APPENDIX D: POWER SPECTRUM ANALYSIS

In order to find the effective number of sampled optical modes, we calculate the spatial power spectrum of the measured speckle, $S(\mathbf{k})$. Since we measure the 2-D intensity speckle $I(\mathbf{x})$, the power spectrums are calculated as the 2-D Fourier transform of the intensity speckles,

$$S(\mathbf{k}) = \left|\iint I(\mathbf{x}) e^{-i2\pi \mathbf{k}\mathbf{x}} d\mathbf{x}\right|^2, \qquad (D1)$$

which is the Wiener-Khinchin theorem. Total $N = 3{,}520$ power spectrums are compounded; and then, azimuthally averaged to get $S(|\mathbf{k}|)$, which is expected to exhibit triangular distribution [2].

From the fitted result, we can calculate the effective area $A_k$ that the speckle field spans in the reciprocal space (Fig. 7), which is directly related to the experimental oversampling ratio, $M'/M = A_k^{-1}$. Therefore, we can deduce the $M = 49{,}600$ and corresponding $M$ to $N$ ratio $\gamma = 14$.

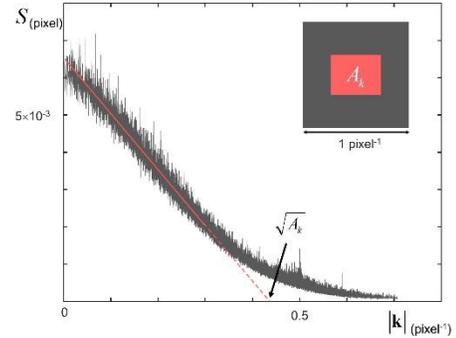

FIG. 7. Power spectrum of speckle. We plot azimuthally averaged power spectrum $S(|\mathbf{k}|)$, which is linearly decreasing (i.e., triangular distribution) as the $|\mathbf{k}|$ increases. We use the 'pixel unit' that indicates the physical pixel size of the used camera as a unit. The inset depicts the effective area $A_k$ of speckle field spans in the reciprocal space. For rectangular windows as in our experiments, the $x$-intercept indicates $\sqrt{A_k} = 0.436$.

## APPENDIX E: VECTOR FIELD PREPARATION

For the pure states, the initial polarization state was prepared to be $|A\rangle = \frac{1}{\sqrt{2}}|H\rangle - \frac{1}{\sqrt{2}}|V\rangle$, or $\frac{1}{\sqrt{2}}\begin{pmatrix}1\\-1\end{pmatrix}$. Since $|A\rangle$ is the fast axis of the quarter-wave plate, the polarization was maintained before the SLM. The SLM modulates $|H\rangle$

polarization, and the reflected polarization become $\frac{1}{\sqrt{2}}\begin{pmatrix} e^{i\varphi_r} \\ 1 \end{pmatrix}$, where $\varphi_r$ denotes the phase retardance as the function of SLM position $|r\rangle$. Note that the sign of $|V\rangle$ is changed due to the reflection geometry of the SLM, $(x, y, z) \rightarrow (x, -y, -z)$. After the quarter-wave plate, the output polarization becomes

$$|s\rangle = e^{i\frac{\varphi_r}{2}} \begin{pmatrix} \cos\left(\frac{\varphi_r}{2} + \frac{\pi}{4}\right) \\ \sin\left(\frac{\varphi_r}{2} + \frac{\pi}{4}\right) \end{pmatrix}. \quad (E1)$$

Allocating different phase values over the SLM active plane, the output polarization state exhibits a position dependency called the 'vector field.'

## APPENDIX F: STOKES PARAMETERS MEASUREMENTS

In order to confirm the unpolarization state generated from a He-Ne laser (HNL050R, Thorlabs Inc.), we measured the Stokes parameter by measuring the intensities in six different polarization states. The polarization of the analyzer was changed using a polarizer (LPVISE100-A; Thorlabs Inc.) and a liquid crystal variable retarder (LCVR; LCC1222-A, Thorlabs Inc.). The measured Stokes parameters were (1, 0.006, -0.004, -0.006), which clearly indicates the unpolarized state.

## APPENDIX G: POTENTIAL FACTORS THAT MAY AFFECT THE REQUIRED OVERSAMPLING RATIO

Notice for optical systems having $g \approx M$, the mean transmittance and the transmission eigenvalue distribution of the optical system may also affect the achievable fidelity.

### A. Mean transmittance, $\bar{T}$

The information transmission capacity of the optical system could be quantified by the dimensionless conductance ($g$) [48], If we utilize the optical modes exceed $g$, we should take the reflective information loss and induced mesoscopic correlation between output speckles addressed in Ref. [48] into accounts. However, please notice our experiments have far larger $g = 1.77 \times 10^8$ than the input ($N = 3,520$) and output ($M = 49,600$) optical modes, so the reflective intensity loss does not induce the loss of optical information in our experimental situations

### B. The condition number of a transmission matrix

Since the SSM can be considered as a nonlinear inversion process, the condition number of the transmission matrix may be an important factor. According to Ref. [48], The eigenvalue ($\tilde{\tau}$) distribution of $\hat{t}^\dagger \hat{t}$ of disordered system having $g \gg M$, follows Marcenko-Pastur (MP) distribution. The maximum and minimum eigenvalues ($\tilde{\tau}_{max}$ and $\tilde{\tau}_{min}$) of MP distribution are the function of oversampling ratio ($\gamma$), $\tilde{\tau}_{max} = \langle\tilde{\tau}\rangle\left(1+\gamma^{-1/2}\right)^2$ and $\tilde{\tau}_{min} = \langle\tilde{\tau}\rangle\left(1-\gamma^{-1/2}\right)^2$, where $\langle\tilde{\tau}\rangle$ is the mean of the eigenvalues. Then, the condition number of transmission matrix also becomes the function of $\gamma$,

$$\text{condition number} = \sqrt{\frac{\tilde{\tau}_{max}}{\tilde{\tau}_{min}}} = \frac{\sqrt{\gamma}+1}{\sqrt{\gamma}-1} \quad (G1)$$

Notice the condition number decreases as $\gamma$ increases. Therefore, we can infer that previous numerical results (Fig. 5) already includes the effect of condition number, since the effect of $\gamma$ is already taken into account.

Fig. 8 shows experimental transmission eigenvalue distribution and the MP distribution for $\gamma = 14$. Although the measured condition number (8.00) is far larger than the expected value (1.73), we find the overall trend of eigenvalue distribution follows the MP distribution. We suspect the unexpected tails on the measured eigenvalue distribution is originated from the slight non-zero correlations between the preset planewaves $|p\rangle$, due to the unmodulated portion of used SLM.

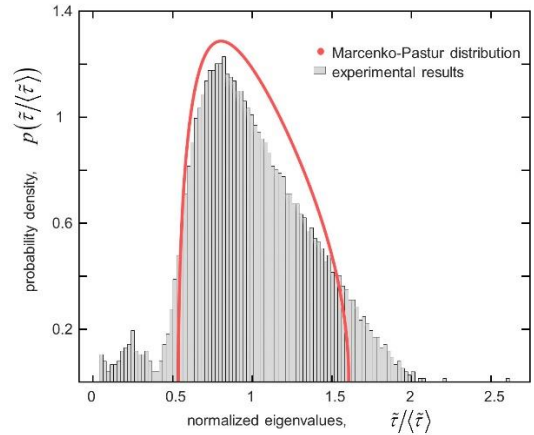

FIG. 8. The histogram of the experimental eigenvalue ($\tilde{\tau}$) distribution of $\hat{t}^\dagger \hat{t}$. Expected Marcenko-Pastur distribution for $\gamma = 14$ is also shown (red line).

## REFERENCES

[1]  J. C. Dainty, *Laser speckle and related phenomena* (Springer-Verlag, 1975).
[2]  J. W. Goodman, *Speckle Phenomena in Optics: Theory and Applications* (Roberts & Company, 2007).
[3]  N. George, C. R. Christensen, J. S. Bennett, and B. D. Guenther, J. Opt. Soc. Am. **66**, 1282 (1976).
[4]  A. Kozma and C. R. Christensen, J. Opt. Soc. Am. **66**, 1257 (1976).
[5]  J. M. Artigas and A. Felipe, J. Opt. Soc. Am. A **5**, 1767 (1988).

[6] J. A. Leendertz, Journal of Physics E: Scientific Instruments **3**, 214 (1970).
[7] D. B. Barker and M. E. Fourney, Opt. Lett. **1**, 135 (1977).
[8] P. V. Farrell and D. L. Hofeldt, Appl. Opt. **23**, 1055 (1984).
[9] R. Erf, *Speckle Metrology* (Elsevier Science, 2012).
[10] B. Redding, S. F. Liew, R. Sarma, and H. Cao, Nature Photonics **7**, 746 (2013).
[11] V. Trivedi, S. Mahajan, V. Chhaniwal, Z. Zalevsky, B. Javidi, and A. Anand, Sensors and Actuators A: Physical **216**, 312 (2014).
[12] K. Kim, H. Yu, J. Koh, J. H. Shin, W. Lee, and Y. Park, Opt. Lett. **41**, 1837 (2016).
[13] S. Popoff, G. Lerosey, M. Fink, A. C. Boccara, and S. Gigan, Nature communications **1**, 81 (2010).
[14] S. M. Popoff, G. Lerosey, R. Carminati, M. Fink, A. C. Boccara, and S. Gigan, Physical Review Letters **104**, 100601 (2010).
[15] Y. Bromberg and H. Cao, Physical Review Letters **112**, 213904 (2014).
[16] A. J. F. Siegert and M. I. o. T. R. Laboratory, *On the Fluctuations in Signals Returned by Many Independently Moving Scatterers* (Radiation Laboratory, Massachusetts Institute of Technology, 1943).
[17] D. A. Boas and A. K. Dunn,  (SPIE, 2010), p. 12.
[18] K. Kim, H. Yu, K. Lee, and Y. Park, Scientific Reports **7**, 44435 (2017).
[19] K. Lee and Y. Park, Nature Communications **7**, 13359 (2016).
[20] Y. Baek, K. Lee, and Y. Park, arXiv preprint arXiv:1802.10321 (2018).
[21] H. Kwon, E. Arbabi, S. M. Kamali, M. Faraji-Dana, and A. Faraon, Optica **5**, 924 (2018).
[22] T. Čižmár and K. Dholakia, Nature Communications **3**, 1027 (2012).
[23] Y. Choi, C. Yoon, M. Kim, T. D. Yang, C. Fang-Yen, R. R. Dasari, K. J. Lee, and W. Choi, Physical Review Letters **109**, 203901 (2012).
[24] L. Mandel, E. Wolf, and C. U. Press, *Optical Coherence and Quantum Optics* (Cambridge University Press, 1995).
[25] B. E. A. Saleh and M. C. Teich, *Fundamentals of Photonics* (Wiley, 2013).
[26] E. Jakeman and P. N. Pusey, Journal of Physics A: Mathematical and General **8**, 369 (1975).
[27] E. Jakeman and J. G. McWhirter, Applied Physics B **26**, 125 (1981).
[28] P. A. Mello, E. Akkermans, and B. Shapiro, Physical Review Letters **61**, 459 (1988).
[29] A. P. Mosk, A. Lagendijk, G. Lerosey, and M. Fink, Nature Photonics **6**, 283 (2012).
[30] G. C. Wick, Physical Review **80**, 268 (1950).
[31] U. Leonhardt, *Measuring the Quantum State of Light* (Cambridge University Press, 1997).
[32] R. Martínez-Herrero, P. M. Mejías, and G. Piquero, *Characterization of Partially Polarized Light Fields* (Springer Berlin Heidelberg, 2009).
[33] F. Gori, M. Santarsiero, and R. Borghi, Opt. Lett. **31**, 858 (2006).
[34] K. H. Kagalwala, G. Di Giuseppe, A. F. Abouraddy, and B. E. A. Saleh, Nature Photonics **7**, 72 (2012).
[35] A. F. Abouraddy, K. H. Kagalwala, and B. E. A. Saleh, Opt. Lett. **39**, 2411 (2014).
[36] D. Bicout, C. Brosseau, A. S. Martinez, and J. M. Schmitt, Physical Review E **49**, 1767 (1994).
[37] M. Xu and R. R. Alfano, Physical Review Letters **95**, 213901 (2005).
[38] I. M. Vellekoop and A. P. Mosk, Opt. Lett. **32**, 2309 (2007).
[39] J.-H. Park, C. Park, H. Yu, Y.-H. Cho, and Y. Park, Opt. Lett. **37**, 3261 (2012).
[40] J.-H. Park, C. Park, H. Yu, Y.-H. Cho, and Y. Park, Opt. Express **20**, 17010 (2012).
[41] O. N. Dorokhov, Solid State Communications **51**, 381 (1984).
[42] Y. Imry, EPL (Europhysics Letters) **1**, 249 (1986).
[43] Y. V. Nazarov, Physical Review Letters **73**, 134 (1994).
[44] M. C. W. van Rossum and T. M. Nieuwenhuizen, Reviews of Modern Physics **71**, 313 (1999).
[45] S. Rotter and S. Gigan, Reviews of Modern Physics **89**, 015005 (2017).
[46] R. Horisaki, R. Egami, and J. Tanida, Opt. Express **24**, 3765 (2016).
[47] J. R. DeVore, J. Opt. Soc. Am. **41**, 416 (1951).
[48] C. W. Hsu, S. F. Liew, A. Goetschy, H. Cao, and A. Douglas Stone, Nature Physics **13**, 497 (2017).
[49] J. Yoon, K. Lee, J. Park, and Y. Park, Opt. Express **23**, 10158 (2015).
[50] Q. Zhan, Adv. Opt. Photon. **1**, 1 (2009).
[51] K. Lee and Y. Park, MATLAB code, figshare (2019) [retrieved 13 March, 2019], 10.6084/m9.figshare.7835516.v2
[52] R. Jozsa, Journal of Modern Optics **41**, 2315 (1994).
[53] J. Ellis and A. Dogariu, Physical Review Letters **95**, 203905 (2005).
[54] R. F. Wagner, S. W. Smith, J. M. Sandrik, and H. Lopez, IEEE Transactions on Sonics and Ultrasonics **30**, 156 (1983).
[55] J. Wang and A. Z. Genack, Nature **471**, 345 (2011).
[56] M. Sutton, S. G. J. Mochrie, T. Greytak, S. E. Nagler, L. E. Berman, G. A. Held, and G. B. Stephenson, Nature **352**, 608 (1991).
[57] M. Jang *et al.*, Nature Photonics **12**, 84 (2018).
[58] J. Park, K. Lee, and Y. Park, Nature Communications **10**, 1304 (2019).